\documentstyle[12pt,amssymb]{article}
\newcommand{\g}{\mathcal{G}}

\def\beq{\begin{equation}}
\def\eeq{\end{equation}}
\def\bea{\begin{eqnarray}}
\def\eea{\end{eqnarray}}
\def\bq{\begin{quote}}
\def\eq{\end{quote}}

\def\gappeq{\mathrel{\rlap {\raise.5ex\hbox{$>$}}
{\lower.5ex\hbox{$\sim$}}}}

\def\lappeq{\mathrel{\rlap{\raise.5ex\hbox{$<$}}
{\lower.5ex\hbox{$\sim$}}}}

\def\Toprel#1\over#2{\mathrel{\mathop{#2}\limits^{#1}}}
\begin {document}

\pagestyle{empty}
\begin{flushright}
{CERN-TH/2001-008}\\
\end{flushright}
\vspace*{5mm}
\begin{center}
{\bf Spinors, superalgebras and the signature of space-time} \\
\vspace*{1cm} 
{\bf S. Ferrara} \\
\vspace{0.3cm}

Theoretical Physics Division, CERN \\
CH - 1211 Geneva 23 \\
\vspace*{3.5cm}  
{\bf ABSTRACT} \\ \end{center}
\vspace*{5mm}
\noindent
Superconformal algebras embedding space-time in any dimension and signature are
considered.  Different real forms of the $R$-symmetries arise both for usual
space-time signature (one time) and for Euclidean or exotic signatures
(more than one times).  Application of these superalgebras are found in the
context of supergravities with 32 supersymmetries, in any dimension $D \leq
11$.  These theories are related to $D = 11, M, M^*$ and $M^\prime$ theories or
$D = 10$, IIB, IIB$^*$ theories when compactified on Lorentzian tori.  All
dimensionally reduced theories fall in three distinct phases specified by the
number of (128 bosonic) positive and negative norm states:  $(n^+,n^-) =
(128,0), (64,64), (72,56)$.
\vspace*{1.5cm}
\begin{center} 
{\it Based on talks given at the \\
``XIV Sigrav  Congress on Recents Developments
in General Relativity",\\ Genoa, Italy, September 2000\\
Workshop on ``30 years of Supersymmetry", Minneapolis, U.S.A., October 2000\\
``Dirac Anniversary Meeting", ICTP, Trieste, Italy, November 2000 }
\end{center}
\vspace*{0.5cm}

\begin{flushleft} 
CERN-TH/2001-008 \\
January 2001
\end{flushleft}
\vfill\eject

\setcounter{page}{1}
\pagestyle{plain}

\section{Introduction}

We describe superconformal algebras embedding space-time $V_{s,t}$ with
arbitrary dimension $D = s + t$ and signature $\rho = s - t$ \cite{aaa}.
The relation between $R$-symmetries and space-time signatures is
elucidated \cite{bb}.  For
supergravities with 32 supersymmetries, there exist theories in three distinct
phases specified by the number of (128 bosonic) positive and negative metric
``states":  $(n^+,n^-) = (128,0), (64,64), (72,56)$.  This fact is closely
related to the compactification of $M, M^*$ and $M^\prime$ theories on Lorentzian
tori \cite{c2}.

When dimensionally reduced to $D = 3$, two of these phases are Minkowskian with
$\sigma$-model $E_{8(8)}/SO(16)$ and $E_{8(8)}/SO(8,8)$;  one is Euclidean with
$\sigma$-model $E_{8(8)}/SO^*(16)$.  The three real forms of $D_8$ correspond to the
superconformal algebras $Osp(16/4,R), Osp(8,8/4,R)$ and $Osp(16^*/2,2)$ associated
to real and quaternionic spinors respectively.

The paper is organized as follows:

In Sections 2 and 3, we describe symmetry and reality properties of spinors in
arbitrary space-time signatures and dimensions.  In Sections 4 and 5, we
consider real forms of classical Lie algebras and superalgebras.  In Section 6,
we study extended superalgebras.  Supersymmetry algebras with non-compact
$R$-symmetries arise in the context of Euclidean theories as well as in theories
with exotic (more than one times) signature.  In Section 7, we show that
non-compact $R$-symmetries arise when $D = 11$  and $M, M^*, M^\prime$ theories
introduced by C.~Hull \cite{c2} are compactified to lower dimensions on Lorentzian tori.

Euclidean supergravity from time-reduction \cite{cllpst} of $M$-theory was considered by Hull
and Julia \cite{cj} and BPS branes in theories with more than one times were studied by
 Hull and Khuri \cite{ck}. 
Hull's observation is that theories on exotic space-time signatures arise by
$T$-duality on time-direction from conventional $M$ theory on string theory so this can
perhaps be regarded as different phases of the same non-perturbative $M$-theory \cite{c2}.
Irrespectively of whether it is sensible to use time-like $T$-duality in string
theory \cite{gm}, supergravity theories in exotic space-time dimensions certainly exist, on
the basis of invariance principles due to the existence of appropriate
superalgebras \cite{aaa, bb}.
 $U$-duality also implies the existence of two new kinds of type II string
 theories, IIA$^*$, II$^*$B in $V_{9,1}$ space-time where all bosonic $RR$ fields
 have reversed sign for the metric \cite{c1}.  A strong evidence of this reasoning is that
 the $U$-duality groups in all lower dimensions are the same, independently of
 the space-time signature.
 BPS states are classified by orbits of the
 $U$-duality group.  These orbits were given in Refs.~\cite{fg,fm} and \cite{lps}.  Since 
 they 
 depend only on the $U$-duality group and not on the $R$-symmetry, it follows
 that the orbit classification is insensitive to the signature of space-time and
 is still valid for $M^*, M^\prime$ theories.
 
 Finally, in Section 8 we describe state counting of supergravities in exotic
 space-time.  This is given in $D = 4$ and $D = 3$ dimensions, as well as in the
 $D = 11, 10$ original dimensions.  This counting shows that there are
 essentially three distinct phases, in any dimension and signature where the
 (128) bosonic degrees of freedom fall in positive and negative norm classes
 $(n^+,n^-) = (128,0), (64,64), (72,56)$.  The first two are Minkowskian phases
 (one time or more than one times), the latter is the Euclidean phase (no time).  State
 counting in theories with 16 supersymmetries is also given, and again three different
 phases emerge.

\section{Properties of spinors of SO($\mathbf{V}$)\label{seclif}}

Let $V$ be a real vector space of dimension $D=s+t$ and
$\{v_\mu\}$ a basis of it. On $V$ there is a non degenerate
symmetric bilinear form, which in the basis is given by the matrix
 $$\eta_{\mu\nu}={\rm diag}(+,\dots (s \;{\rm times})\dots,
+,-,\dots (t \;{\rm times})\dots, -).$$

We consider the group Spin($V$), the unique double covering of the
connected component of ${\rm SO}(s,t)$ and its spinor
representations. A spinor representation of Spin$(V)^{\Bbb C}$ is an
irreducible complex representation whose highest weights are the
fundamental weights corresponding to the right  extreme nodes in
the Dynkin diagram. These do not descend to representations of
SO$(V)$. A spinor-type representation is any irreducible
representation that does not descend to SO$(V)$. A spinor
representation of Spin$(V)$ over the reals  is an irreducible
representation over the reals whose complexification is a direct
sum of spin representations \cite{ch}--\cite{de}.

 Two parameters, the signature $\rho$
mod(8) and the dimension $D$ mod(8) classify the properties of the
 spinor representation. Through this paper we will use the
following notation:
 
 $$ \rho=s-t=\rho_0 +8n,\qquad D=s+t=D_0 +8p,$$
 
where $\rho_0, D_0= 0,~\dots, 7$. We set $m=p-n,$ so
\begin{eqnarray*}s&=&\frac{1}{2}(D +\rho)=\frac{1}{2}(\rho_0
+D_0)+8n+4m,\\
t&=&\frac{1}{2}(D-\rho)=\frac{1}{2}(D_0-\rho_0)+4m.\end{eqnarray*}

The signature  $\rho$ mod(8) determines if the spinor
representations are of the real($\Bbb R$), quaternionic  ($\Bbb H$) or complex
($\Bbb C$) type.  Also note that reality properties depend only on $|\rho|$ since
Spin$(s,t)$ = Spin$(t,s)$.

The dimension-$D$ mod(8) determines the nature of the
Spin($V$)-morphisms of the spinor representation $S$. Let $g\in
{\rm Spin}(V)$ and let  $\Sigma(g):S\longrightarrow S$  and
$L(g):V\longrightarrow V$ the spinor and vector representations of
$l\in{\rm Spin}(V)$ respectively. Then a
 map $A$
$$ A: S\otimes S
\longrightarrow \Lambda^k ,$$
 where $\Lambda^k=\Lambda^k(V)$ are
the $k$-forms on $V$, is a Spin($V$)-morphism if
$$A(\Sigma(g)s_1\otimes \Sigma(g)s_2)=L^k(g)A(s_1\otimes s_2). $$

\medskip

In Tables \ref{realprop} and \ref{morphisms}, reality and symmetry properties of spinors are reported.

\begin{table}[ht]
\begin{center}
\begin{tabular} {|c |c| c||c|c|c|}
\hline $\rho_0$(odd) &Real dim($S$) & Reality &$\rho_0$(even)
&Real dim($S^{\pm}$) & Reality\\ \hline \hline 1& $2^{(D-1)/2}$
&$\Bbb R$ & 0 &$2^{D/2-1}$&$\Bbb R$ \\ \hline 3& $2^{(D+1)/2}$ &$\Bbb H$& 2
&$2^{D/2}$&$\Bbb C$ \\ \hline 5& $2^{(D+1)/2}$ &$\Bbb H$& 4
&$2^{D/2}$&$\Bbb H$\\ \hline 7& $2^{(D-1)/2}$ &$\Bbb R$ & 6
&$2^{D/2}$&$\Bbb C$ \\ \hline
\end{tabular}
\caption{Reality properties of spinors}\label{realprop}
\end{center}
\end{table}

\bigskip
 
 \begin{table}[ht]
\begin{center}
\begin{tabular} {|c |c|c||c|c|}
\hline \multicolumn{1}{|c|}{$D$} &\multicolumn{2}{| c||}{$k $
even}&\multicolumn{2}{|c|}{$k $ odd}\\\hline\hline & Morphism &
Symmetry &Morphism&Symmetry\\
  \hline
  0& $S^\pm\otimes S^\pm\rightarrow\Lambda^k$&$(-1)^{k(k-1)/2}$& $S^\pm\otimes
  S^\mp\rightarrow\Lambda^k$&\\\hline
  1& $S\otimes S\rightarrow\Lambda^k$&$(-1)^{k(k-1)/2}$& $S\otimes
  S\rightarrow\Lambda^k$&$(-1)^{k(k-1)/2}$\\\hline
  2& $S^\pm\otimes S^\mp\rightarrow\Lambda^k$&& $S^\pm\otimes
  S^\pm\rightarrow\Lambda^k$&$(-1)^{k(k-1)/2}$\\\hline
  3& $S\otimes S\rightarrow\Lambda^k$&$-(-1)^{k(k-1)/2}$& $S\otimes
  S\rightarrow\Lambda^k$&$(-1)^{k(k-1)/2}$\\\hline
  4& $S^\pm\otimes S^\pm\rightarrow\Lambda^k$&$-(-1)^{k(k-1)/2}$& $S^\pm\otimes
  S^\mp\rightarrow\Lambda^k$&\\\hline
  5& $S\otimes S\rightarrow\Lambda^k$&$-(-1)^{k(k-1)/2}$& $S\otimes
  S\rightarrow\Lambda^k$&$-(-1)^{k(k-1)/2}$\\\hline
  6& $S^\pm\otimes S^\mp\rightarrow\Lambda^k$&& $S^\pm\otimes
  S^\pm\rightarrow\Lambda^k$&$-(-1)^{k(k-1)/2}$\\\hline
  7& $S\otimes S\rightarrow\Lambda^k$&$(-1)^{k(k-1)/2}$& $S\otimes
  S\rightarrow\Lambda^k$&$-(-1)^{k(k-1)/2}$\\\hline

\end{tabular}
\caption{Properties of morphisms}\label{morphisms}
\end{center}
\end{table}

\section{Orthogonal, symplectic and linear spinors}

We now consider morphisms $$S\otimes S\longrightarrow
\Lambda^0\simeq \Bbb C.$$ If a morphism of this kind exists, it is
unique up to a multiplicative factor. The vector space of the
spinor representation then has a bilinear form invariant under
Spin($V$).
 Looking at Table \ref{morphisms}, one can see that this morphism
 exists except for $D_0=2,6$, where instead a morphism
 $$S^{\pm}\otimes S^{\mp}\longrightarrow \Bbb C$$
 occurs.

 We shall call a spinor representation orthogonal if it has a
 symmetric, invariant bilinear form. This happens for $D_0=0,1,7$ and
 Spin$(V)^{\Bbb C}$ (complexification of Spin($V$)) is then  a subgroup
 of the complex orthogonal group ${\rm SO}(n,\Bbb C)$, where $n$ is the
 dimension of the spinor representation (Weyl spinors for $D$ even).
 The generators
 of SO$(n,\Bbb C)$ are  $n\times n$ antisymmetric matrices. These are obtained
 in terms of the morphisms
$$S\otimes S\longrightarrow\Lambda^k,$$ which are  antisymmetric.
This gives  the decomposition of the adjoint representation of
${\rm SO}(n,\Bbb C)$ under the subgroup ${\rm Spin}(V)^{\Bbb C}$.
 In particular, for $k=2$ one obtains the generators of ${\rm Spin}(V)^{\Bbb C}$.

 A spinor representation is called symplectic if it has an
 antisymmetric, invariant bilinear form. This is the case for
 $D_0=3,4,5$. ${\rm Spin}(V)^{\Bbb C}$ is a subgroup of the symplectic group
 ${\rm Sp}(2p,\Bbb C)$, where $2p$ is the dimension of the
 spinor representation. The Lie algebra ${\rm sp}(2p,\Bbb C)$ is formed by all the
 symmetric matrices, so it is  given in terms of the
  morphisms $S\otimes S\rightarrow \Lambda^k$, which are symmetric.
 The generators
 of ${\rm Spin}(V)^{\Bbb C}$ correspond to $k=2$ and  are symmetric matrices.

 For $D_0=2,6$
 one has an
 invariant morphism
 $$B:S^{+}\otimes S^{-}\longrightarrow \Bbb C.$$
One of the representations $S^+$ and $S^-$ is
one the contragradient (or dual) of the other.
 The spin representations extend to  representations of the
 linear group ${\rm GL}(n,\Bbb C)$, which leaves the pairing  $B$ invariant. These
 spinors are called linear. Spin($V)^{\Bbb C}$ is a subgroup of the simple factor 
 SL$(n,\Bbb C)$.

These properties depend exclusively on the dimension \cite{de}. When combined with the
reality properties, which depend on
$\rho$, one obtains real groups embedded in ${\rm SO}(n,\Bbb C)$, ${\rm
Sp}(2p,\Bbb C)$ and
${\rm GL}(n,\Bbb C)$, which have an action on the space of the real spinor
representation $S^\sigma$. The real groups contain ${\rm Spin}(V)$ as a subgroup.

We first need some general facts about real forms of simple Lie
algebras \cite{de}. Let $S$ be a complex vector space of dimension $n$, which
carries an irreducible representation of a complex Lie algebra
$\g$. Let $G$ be the complex Lie group associated to $\g$.  Let
$\sigma$ be a conjugation or a  pseudoconjugation on $S$ such that
$\sigma X\sigma^{-1}\in \g$ for all $X \in \g.$ Then  the map
$$X\mapsto X^\sigma=\sigma X\sigma^{-1}$$ is a conjugation of
$\g$. We shall write 
$${\g}^\sigma=\{X\in \g|X^\sigma=X\}.$$
${\g}^\sigma$ is a real form of $\g$; if $\tau=h\sigma h^{-1}$,
with $h\in \g$, ${\g}^\tau=h{\g}^\sigma h^{-1}$. We have ${\g}^\sigma={\g}^{\sigma'}$
if and only if $\sigma'=\epsilon \sigma$, for $\epsilon$ a scalar with $|\epsilon|=1$;
in particular, if ${\g}^\sigma$ and ${\g}^\tau$ are conjugate by $G$,
$\sigma$ and $\tau$ are both conjugations or both pseudoconjugations. The conjugation can
also be defined on the group $G$, $g\mapsto \sigma g\sigma^{-1}$.

\section{Real forms of the classical Lie algebras}

We  describe  the real forms of the classical Lie algebras from this
point of view \cite{aaa}. (See also Ref.~\cite{he}.)
\vfill\eject
\paragraph{Linear algebra, sl($\mathbf{ S}$).}

\subparagraph{(a)} If $\sigma$ is a conjugation of $S$, then
there is an isomorphism  $S\rightarrow  {\Bbb C}^n$  such that $\sigma$
goes over to the standard conjugation of ${\Bbb C}^n$. Then
${\g}^\sigma\simeq{\rm sl}(n,\Bbb R)$. (The conjugation acting  on gl$(n,\Bbb C)$ gives
 the real form gl($n,\Bbb R$)).

\subparagraph{(b)} If $\sigma$ is a pseudoconjugation and $\g$
does not leave invariant a non-degenerate bilinear form, then there
is an isomorphism of $S$ with ${\Bbb C}^n$, $n=2p$ such that $\sigma$
goes over to $$(z_1,\dots,z_p,z_{p+1},\dots, z_{2p})\mapsto
(z^*_{p+1},\dots, z^*_{2p},-z^*_1,\dots,-z^*_p).$$ Then
${\g}^\sigma\simeq {\rm su}^*(2p)$. (The pseudoconjugation acting in
on gl$(2p,\Bbb C)$ gives
 the real form ${\rm su}^*(2p)\oplus{\rm so}(1,1)$.)

To see this, it is sufficient to prove that ${\g}^\sigma$ does not leave
invariant any non-degenerate hermitian form, so it cannot be of
the type su$(p,q)$. Suppose that $\langle\cdot ,\cdot\rangle$ is a
${\g}^\sigma$-invariant, non-degenerate hermitian form.  Define
$(s_1,s_2):=\langle\sigma (s_1),s_2\rangle$. Then $(\cdot ,\cdot)$
is bilinear and ${\g}^\sigma$-invariant, so it is also
$\g$-invariant.

\subparagraph{(c)} The remaining cases, following E. Cartan's
classification of real forms of simple Lie algebras, are
${\rm su}(p,q)$, where a non-degenerate hermitian bilinear form is
left invariant. They do not correspond to a conjugation or
pseudoconjugation on $S$, the space of the fundamental
representation. (The real form of gl$(n,\Bbb C)$ is in this case u$(p,q)$).

\paragraph{Orthogonal algebra, so($\mathbf{S}$).} $\g$ leaves invariant a non-degenerate,
symmetric bilinear form. We will denote it by $(\cdot,\cdot)$.

\subparagraph{(a)} If $\sigma$ is a conjugation preserving $\g$, one can prove that
 there is an isomorphism of
$S$ with ${\Bbb C}^n$ such that  $(\cdot,\cdot)$ goes to the standard
form and ${\g}^\sigma$ to ${\rm so}(p,q)$, $p+q=n$. Moreover, all
${\rm so}(p,q)$ are obtained in this form.

\subparagraph{(b)} If $\sigma$ is a pseudoconjugation preserving $\g$,
 ${\g}^\sigma$ cannot be
any of the ${\rm so}(p,q)$. By Cartan's classification, the only
other possibility is that ${\g}^\sigma\simeq {\rm so}^*(2p)$. There
is an isomorphism of $S$ with ${\Bbb C}^{2p}$ such that $\sigma$ goes to
$$(z_1,\dots, z_p,z_{p+1},\dots, z_{2p})\mapsto (z^*_{p+1},\dots,
z^*_{2p},-z^*_{1},\dots, -z^*_{p}).$$

\paragraph{Symplectic algebra, sp($\mathbf{S}$).} We denote by $(\cdot,\cdot)$ the symplectic
form on $S$. \subparagraph{(a)}If $\sigma$ is a conjugation
preserving $\g$, it is clear  that there is an isomorphism of $S$
with ${\Bbb C}^{2p}$, such that ${\g}^\sigma\simeq {\rm sp}(2p,\Bbb R)$.

\subparagraph{(b)}If $\sigma$ is a pseudoconjugation preserving
$\g$, then ${\g}^\sigma\simeq {\rm usp}(p,q)$, $p+q=n=2m, \; p=2p',\;
q=2q' $. All the real forms ${\rm usp}(p,q)$ arise in this way.
There is an isomorphism of $S$ with ${\Bbb C}^{2p}$ such that $\sigma$
goes to $$(z_1,\dots z_m,z_{m+1},\dots z_{n})\mapsto
J_mK_{p',q'}(z^*_1,\dots z^*_m,z^*_{m+1},\dots z^*_{n}),$$ where

$$
J_m=\pmatrix {0 & I_{m\times m}\cr
-I_{m\times m} & 0},\qquad
 K_{p',q'}=\pmatrix {-I_{p'\times p'}
&0&0&0\cr
0&I_{q'\times q'}&0&0\cr 
0&0&-I_{p'\times
p'}&0\cr
0&0&0&I_{q'\times q'}}.
$$

\bigskip

 In Section 2 we saw that there is  a conjugation on $S$
when  the spinors are real and a pseudoconjugation when they are
quaternionic \cite{aaa} (both denoted by $\sigma$). We have a group,
${\rm SO}(n,\Bbb C)$, ${\rm Sp}(2p,\Bbb C)$ or
 ${\rm GL}(n,\Bbb C)$ acting on $S$ and  containing ${\rm Spin}(V)^{\Bbb C}$.  We note
that this group is minimal in the classical group series. If the
Lie algebra $\g$ of this group is stable under the conjugation
$$X\mapsto \sigma X\sigma^{-1}~,$$ 
then the real Lie algebra
${\g}^\sigma$ acts on $S^\sigma$ and contains the Lie algebra of
${\rm Spin}(V)$. We shall call it the Spin($V$) algebra.

Let $B$ be the space of ${\rm Spin}(V)^{\Bbb C}$-invariant bilinear forms
on $S$. Since the representation on $S$ is irreducible, this space
is at most one-dimensional. Let it be one-dimensional,  let
 $\sigma$ be  a conjugation or a
pseudoconjugation, and let $\psi\in B$. We define a conjugation
in the space $B$ as \begin{eqnarray*} B&\longrightarrow &B\\
\psi&\mapsto& \psi^\sigma\end{eqnarray*} $$
\psi^\sigma(v,u)=\psi(\sigma(v),\sigma(u))^*.$$ It is then
immediate that we can choose $\psi\in B$ such that
$\psi^\sigma=\psi$. Then if $X$ belongs to the Lie algebra
preserving $\psi$, so does $\sigma X\sigma^{-1}$.

One can determine   the real Lie algebras  in each
case \cite{aaa}. All the possible cases must be studied separately. 
 All dimension and signature relations are
mod(8). In the following, a relation like ${\rm Spin}(V)\subseteq
G$ for a group $G$ will mean  that the image of ${\rm Spin}(V)$
under the spinor representation  is in the connected component of
$G$. The same applies for the  relation ${\rm Spin}(V)\simeq G$.  For $\rho$ =
0, 1, 7 spin algebras commute with a conjugation, for $\rho$ = 3, 4, 5 they commute
with a pseudoconjugation.  For $\rho$ = 2, 6 they are complex.  The complete
classification is reported in Table 3.

\begin{table}[ht]
\begin{center}
\begin{tabular} {|l |l|l|}
\hline 
Orthogonal&Real, $\rho_0=1,7$&${\rm so}(2^{\frac{(D-1)}{2}},\Bbb R)$ if $D=\rho$\\
\cline{3-3}
$D_0=1,7$&& ${\rm so}(2^{\frac{(D-1)}{2}-1},2^{\frac{(D-1)}{2}-1})$
if $D\neq\rho$\\
\cline{2-3} & Quaternionic,~$\rho_0=3,5$&${\rm so}^*(2^{\frac{(D-1)}{2}})$\\
\hline\hline 
Symplectic&Real,
$\rho_0=1,7$& ${\rm sp}(2^{\frac{(D-1)}{2}},\Bbb R)$\\
\cline{2-3}
$D_0=3,5$& Quaternionic,~$\rho_0=3,5$& ${\rm
usp}(2^{\frac{(D-1)}{2}},\Bbb R)$  if $D=\rho$\\
\cline{3-3} && ${\rm
usp}(2^{\frac{(D-1)}{2}-1},2^{\frac{(D-1)}{2}-1})$ if $D\neq
\rho$\\
\hline\hline\hline
 Orthogonal&Real, $\rho_0=0$&${\rm
so}(2^{\frac{D}{2}-1},\Bbb R)$ if $D=\rho$\\  \cline{3-3} $D_0=0$&&
${\rm so}(2^{\frac{D}{2}-2},2^{\frac{D}{2}-2})$
 if
$D\neq\rho$\\\cline{2-3}
 & Quaternionic,~$\rho_0=4$&${\rm
so}^*(2^{\frac{D}{2}-1})$\\\cline{2-3} &Complex, $\rho_0=2,6$&
${\rm so}(2^{\frac{D}{2}-1},{\Bbb C})_{\Bbb R}$\\
\hline\hline Symplectic&Real,
$\rho_0=0$& ${\rm sp}(2^{\frac{D}{2}-1},\Bbb R)$\\\cline{2-3} $D_0=4$&
Quaternionic,~$\rho_0=4$& ${\rm usp}(2^{\frac{D}{2}-1},\Bbb R)$ if
$D=\rho$\\
\cline{3-3} && ${\rm
usp}(2^{\frac{D}{2}-2},2^{\frac{D}{2}-2})$ if $D\neq
\rho$\\\cline{2-3} &Complex, $\rho_0=2,6$&${\rm
sp}(2^{\frac{D}{2}-1},{\Bbb C})_{\Bbb R}$\\
\hline\hline
Linear&Real,
$\rho_0=0$& ${\rm sl}(2^{\frac{D}{2}-1},\Bbb R)$\\
\cline{2-3}
$D_0=2,6$& Quaternionic,~$\rho_0=4$& ${\rm
su}^*(2^{\frac{D}{2}-1})$ \\\cline{2-3} &Complex, $\rho_0=2,6$&
${\rm su}(2^{\frac{D}{2}-1})$ if $D= \rho$\\
\cline{3-3} &&${\rm
su}(2^{\frac{D}{2}-2},2^{\frac{D}{2}-2})$ if $D\neq \rho$\\\hline
\end{tabular}
\caption{Spin$(s,t)$ algebras}\label{spinalgebra}
\end{center}
\end{table}

\section{ Spin$\mathbf{(V)}$ superalgebras}

 We now consider the embedding of ${\rm Spin}(V)$ in simple real superalgebras.
We require in general  that the odd generators be in a real
  spinor representation of
  ${\rm Spin}(V)$. In the cases $D_0=2,6$, $\rho_0=0,4$ we have to allow the two independent
irreducible  representations $S^+$ and $S^-$ in the superalgebra, since  the
relevant morphism is
$$S^+\otimes S^-\longrightarrow \Lambda^2.$$
The algebra is then non-chiral.

We first consider minimal superalgebras \cite{na,vp}, i.e. those with the
  minimal even subalgebra. From the classification of simple superalgebras
  \cite{nrs}--\cite{mp} one
obtains the results listed in Table~4.

\begin{table}
\begin{center}
\begin{tabular} {|c|c|l|l|}
\hline
 $D_0$   & $\rho_0$& Spin($V$) algebra&Spin($V$) superalgebra\\\hline\hline
 1,7& 1,7& so($2^{(D-3)/2},2^{(D-3)/2}$)&  \\\hline
 1,7& 3,5 & so$^*$($2^{(D-1)/2}$)&osp$(2^{(D-1)/2})^*|2)$ \\\hline
 3,5& 1,7& sp($2^{(D-1)/2},\Bbb R$)&osp$(1|2^{(D-1)/2},\Bbb R)$\\\hline
 3,5& 3,5& usp($2^{(D-3)/2},2^{(D-3)/2}$)&
 \\\hline\hline
 0& 0&so($2^{(D-4)/2},2^{(D-4)/2}$) &  \\\hline
 0& 2,6&so($2^{(D-2)/2},{\Bbb C})^{\Bbb R}$ &  \\\hline
 0& 4 & so$^*(2^{(D-2)/2}$)&osp$(2^{(D-2)/2})^*|2)$\\\hline
 2,6&  0&sl$(2^{(D-2)/2},\Bbb R)$&sl$(2^{(D-2)/2}|1)$ \\\hline
 2,6&2,6& su$(2^{(D-4)/2},2^{(D-4)/2})$&su$(2^{(D-4)/2},2^{(D-4)/2}|1)$\\\hline
2,6 & 4 &su$^*(2^{(D-2)/2}))$&su$^*(2^{(D-2)/2})|2)$ \\\hline
  4&0&sp($2^{(D-2)/2},\Bbb R$)&osp$(1|2^{(D-2)/2},\Bbb R)$ \\\hline
 4&2,6&sp($2^{(D-2)/2},{\Bbb C})^{\Bbb R}$&osp$(1|2^{(D-2)/2},\Bbb C)$ \\\hline
 4&4&usp($2^{(D-4)/2},2^{(D-4)/2}$)&  \\\hline
\end{tabular}
\caption{Minimal Spin($V$) superalgebras.}\label{min}
\end{center}
\end{table}

We note that the even part of the minimal superalgebra contains the Spin($V$) algebra
 obtained in Section~4 as a simple factor. For all quaternionic cases,
 $\rho_0=3,4,5$, a second simple factor SU(2) is present. For the linear cases there is
 an additional
non-simple factor, SO(1,1) or U(1), as discussed in Section~4.

For $D=7$ and $\rho=3$ there is actually a smaller
 superalgebra, the exceptional  superalgebra  $f(4)$ with
 bosonic part  spin(5,2)$\times$su(2). The superalgebra
appearing in  Table \ref{min} belongs to the classical series and
its  even part is  so$^*(8)\times$su(2), so$^*(8)$ being the
Spin$(5,2)$-algebra.

Keeping the same number of odd generators, the  maximal simple  superalgebra
 containing ${\rm Spin}(V)$ is  an
 orthosymplectic algebra with ${\rm Spin}(V)\subset {\rm Sp}(2n,\Bbb
 R)$\cite{df}--\cite{mg},  $2n$ being the real
dimension of $S$. The various cases  are displayed in  Table~5. We note that the minimal superalgebra is not a
subalgebra of the maximal one, although it is so for the bosonic
parts.

\begin{table}
\begin{center}
\begin{tabular} {|c|c|l|}
\hline
 $D_0$   & $\rho_0$& Orthosymplectic\\\hline\hline
3,5,& 1,7& osp$(1|2^{(D-1)/2},\Bbb R)$ \\\hline
 1,7& 3,5  & osp$(1|2^{(D+1)/2},\Bbb R)$ \\\hline
  0&4 &osp$(1|2^{D/2},\Bbb R)$  \\\hline 4&0& osp$(1|2^{(D-2)/2},\Bbb R)$
\\\hline
 4&2,6  &osp$(1|2^{D/2},\Bbb R)$ \\\hline
  2,6&  0&osp$(1|2^{D/2},\Bbb R)$ \\\hline
  2,6 & 4 & osp$(1|2^{(D+2)/2},\Bbb R)$\\\hline
 2,6&2,6&osp$(1|2^{D/2},\Bbb R)$ \\\hline
\end{tabular}
\caption{Orthosymplectic Spin($V$) superalgebras}\label{max}
\end{center}
\end{table}

\section{Extended Superalgebras}
The present analysis can be generalized to the case of $N$ copies of the spinor
representation of spin$(s,t)$ algebras \cite{bb}.  By looking at the classification of
classical simple superalgebras \cite{na}--\cite{mp}, we find extensions for all $N$, where the number
of supersymmetries is always even if spinors are quaternionic (because of reality
properties) or orthogonal (because of symmetry properties).

In Table~6 the classification is analogous to the one given in Table~4. 
Super-Poincar\'e algebras can be obtained from the simple superalgebras either by
contraction Spin$(s,t) \rightarrow$ InSpin$(s,t-1)$ or as subalgebras Spin$(s,t)
\rightarrow$ InSpin$(s-1,t-1)$. It is important to observe that the
$R$-symmetry may be non-compact for different signatures of space-time.

All these superalgebras have a space-time symmetry that commutes with the
$R$-symmetry group.  We can further extend these superalgebras to
orthosymplectic superalgebras where these symmetries no longer commute.  The
number of fermionic generators remains unchanged.  The bosonic part is simply a
real symplectic algebra, which contains as a subalgebra the direct sum of the
space-time and $R$-symmetry algebra \cite{bvp}--\cite{pw}.  These extensions are
reported in Table~7.
These superalgebras contain by contraction or as subalgebras, all Poincar\'e
superalgebras in any dimension with all possible ``central charges".

 \begin{table}
\begin{center}
\begin{tabular} {c|c|c|l |l|}

\cline{2-5} &$D_0$&$\rho_0$& R-symmetry&Spin$(s,t)$
superalgebra\\\cline{2-5} & 1,7& 1,7& ${\rm
sp}(2N,\Bbb R)$&${\rm
osp}(2^{\frac{D-3}{2}},2^{\frac{D-3}{2}}|2N,\Bbb R)$\\\cline{2-5}
& 1,7& 3,5& ${\rm usp}(2N-2q,2q)$&${\rm
osp}(2^{\frac{D-1}{2}\,*}|2N-2q,2q)$\\\cline{2-5} & 3,5&
1,7& ${\rm so}(N-q,q)$&${\rm
osp}(N-q,q|2^{\frac{D-1}{2}})$\\\cline{2-5} & 3,5& 3,5&
${\rm so}^*(2N)$&${\rm
osp}(2{N}^*|2^{\frac{D-3}{2}},2^{\frac{D-3}{2}})$\\
\cline{2-5}
\cline{2-5} & 0& 0& ${\rm sp}(2N,\Bbb R)$&${\rm
osp}(2^{\frac{D-4}{2}},2^{\frac{D-4}{2}}|2N)$\\\cline{2-5}
& 0& 2,6& ${\rm sp}(2N,{\Bbb C})_{\Bbb R}$&${\rm
osp}(2^{\frac{D-2}{2}}|2N,{\Bbb C})_{\Bbb R}$\\\cline{2-5} & 0& 4&
${\rm usp}(2N-2q,2q)$&${\rm osp}(2^{\frac{D-2}{2}\,*}|2N-2q,2q)$\\\cline{2-5}
& 2,6& 0& ${\rm sl}(N,\Bbb R)$&${\rm
sl}(2^{\frac{D-2}{2}}|N,\Bbb R)$\\\cline{2-5} &2,6& 2,6& ${\rm
su}(N-q,q)$&${\rm
su}(2^{\frac{D-4}{2}},2^{\frac{D-4}{2}}|N-q,q)$\\
\cline{2-5}
& 2,6& 4& ${\rm su}^*(2N,\Bbb R)$&${\rm
su^*}(2^{\frac{D-2}{2}}|2N)$\\\cline{2-5} & 4& 0& ${\rm
so}(N-q,q)$&${\rm osp}(N-q,q|2^{\frac{D-2}{2}})$\\
\cline{2-5} & 4&
2,6& ${\rm so}(N,{\Bbb C})_{\Bbb R}$&${\rm
osp}(N|2^{\frac{D-2}{2}},{\Bbb C})_{\Bbb R}$\\
\cline{2-5} & 4& 4& ${\rm
so}^*(2N)$&${\rm
osp}(2{N}^*|2^{\frac{D-4}{2}},2^{\frac{D-4}{2}})$\\\cline{2-5}
\end{tabular}
\caption{$N$-extended Spin$(s,t)$ superalgebras}\label{next}
\end{center}
\end{table}
\vfill\eject
\vskip0.5cm
\noindent

\begin{table}
\begin{center}
\begin{tabular}{|c|c|l|}
\hline
$D_0$ & $\rho_0$ & $osp(1/2n,R) \supset sp(2n,R)$\\
\hline
1,7 & 1,7 & $sp(2N \times 2^{\frac{D-1}{2}})$ \\
1,7 & 3,5 & $sp(2N \times 2^{\frac{D-1}{2}})$ \\
3,5 & 1,7 & $sp(N \times 2^{\frac{D-1}{2}})$\\
3,5 & 3,5 & $sp(2N \times 2^{\frac{D-1}{2}})$\\
0 & 0 & $sp(2n \times 2^{\frac{D-2}{2}}$)\\
0 & 2,6 & $sp (2N \times 2^{\frac{D}{2}}$)\\
0 & 4 & $sp(2N \times 2^{\frac{D-2}{2}}$)\\
2,6 & 0 & $sp(N \times 2^{\frac{D}{2}}$)\\
2,6 & 2,6 & $sp(N \times 2^{\frac{D}{2}}$)\\
2,6 & 4 & $sp(2N \times 2^{\frac{D}{2}}$)\\
4 & 0 & $sp(N \times 2^{\frac{D-2}{2}}$)\\
4 & 2,6 & $sp(N \times 2^{\frac{D}{2}}$)\\
4 & 4 & $sp(2N \times 2^{\frac{D-2}{2}})$\\
\hline

\end{tabular}
\caption{Orthosymplectic superalgebras}
\end{center}
\end{table}

\section{11D and 10D on V$_{s,t}$ supergravities revisited}

Theories with $N$ super-Poincar\'e supersymmetries can be obtained as subalgebras
of superconformal algebras with $2N$ spinorial generators.  This is so because a
conformal spinor $S_{s,t}$ of the conformal group $SO(s,t)$ always decomposes
into two Poincar\'e spinors of opposite dimensions:
$$
S_{s,t} \rightarrow s^{1/2}_{s-1,t-1} + s^{-1/2}_{s-1,t-1}~.
$$
The $R$-symmetry is inherited from the simple superalgebra whose odd generators
contain the spin group of the conformal group.  Poincar\'e superalgebras with
at most 32 supersymmetries are obtained as subalgebras of superconformal
algebras with at most 64 charges.  Then the $R$-symmetries of these theories are
read from their superconformal extension.  In particular, although a super-Poincar\'e
graviton exists with at most 32 supersymmetries, a conformal graviton multiplet exists
with at most 64 supercharges \cite{mg}--\cite{fs}.

From Table~6, we can read the $R$-symmetries of super-Poincar\'e algebras by
replacing $(D,\rho)$ with $D-2,\rho)$.

Let us now consider maximal supergravity theories in $D = 11~(N=1)$ and $D = 10$,
IIA, IIB.  For $D = 11$, we know that a real spinor exists for $\rho = \pm1$
mod~8.  This implies the existence of two more theories other than $M$ theory
with signature (9,2) and (6,5).

These theories were introduced by C. Hull and called $M^*$ and $M^\prime$
theories \cite{c2}.  They were discovered on the basis of time-like $T$ duality
of type IIA and IIB string theory, which requires the existence of new kinds of
type II strings called IIA$^*$ and IIB$^*$ string theories \cite{c1}.  The strong coupling
limit of IIA$^*$ theory is $M^*$ theory.

By dimensional reduction of $M, M^*$ and $M^\prime$ theories to $D = 10$, new type
IIA theories with signatures (10,0), (9,1), (8,2), (6,4) and (5,5) are found~\cite{c2}.
Type IIA
(10,0) is IIA$_E$ Euclidean supergravity, $(9,1)^*$ supergravity is type IIA$^*$.  In
the (8,2) (6,4) supergravities, there is a single (16) dimensional complex 
Weyl spinor such that $\psi^*_L = \psi_R$ .

In type IIB, the story is rather different:  (2,0) supergravity has an $R$-symmetry
that can be either SO(2) or SO(1,1) for $\rho =0$ or SO$^*$(2) = SO(2) for $\rho =
4$ (see Table~6 for $ d = 4$).  This leads to the existence of two types of type IIB
theories for $\rho = 0$, (9,1) and (5,5) signature called type IIB (SO(2)) and
IIB$^*$ (SO(1,1)) by C.~Hull~\cite{c1} and one theory for (7,3) signature with SO(2)
$R$-symmetry.  Their respective $10D$ $\sigma$-model is $\frac{SL(2,R)}{SO(2)}$ for
type IIB and $\frac{SL(2,R)}{SO(1,1)}$ for IIB$^*$ theories~\cite{c1,cllpst}.

Most interestingly, in type II$^*$ theories, the sign of the kinetic term in the $RR$
 bosonic fields is reversed
with respect to the $NS$ fields case.  Lifting string theory to
$D = 11~M$ theory, the above implies that in $M^*$ theory the three-form part of the
action has a reversed sign with respect to the $M$ and $M^\prime$ theories case \cite{c2}.  

We have just seen that a non-compact $R$-symmetry arises in IIB$^*$ theory in $D =
10$.  To find other non-compact $R$-symmetries, it is sufficient to compactify $M,
M^*$ and $M^\prime$ theories on Lorentzian tori $T^{(p,q)}$.  A property of
$R$-symmetries is that they must contain, as a subgroup, the orthogonal group
$SO(p,q)$ related to the classical moduli space of a Lorentzian torus
$$
{\cal M}_{T^{(p,q)}} = \frac{GL(p+q,R)}{SO(p,q)}~.
$$
Moreover, the $R$-symmetries must also be a subgroup of the $U$-duality group
$E_{11-D(11-D)}$ and be related to reality properties and dimension of space-time
spinors from Table~6.  This uniquely fixes the non-compact form of the $R$-symmetry
$H_D$.

As illustrative examples, let us consider the $D = 5,4$ and 3 cases for space-time
of all possible signatures.  For $D = 5,4,3, M, M^*$ and $M^\prime$ theories we get
$$
V_{(s,11-s)} \rightarrow V_{(s^\prime,t^\prime)} \times T^{(p,q)} ~~~~~
(s = 10,9,6)~~~~~~~(s^\prime+p = s, ~~ t^\prime + q = 11-s)~.
$$
For $ D = 5, 4, 3$, the $R$-symmetries must be different real forms of the $C_4, A_7$
and $D_8$ Lie algebras, appropriate to spinors of $V_{s^\prime,t^\prime}$, and they
must contain $SO(p,q)$ as subgroup.  The appropriate real forms can be read from
Table~8 and are a consequence of Table~6.

\begin{table}
\begin{center}
\begin{tabular}{|c|c|c|c|l|}
\hline
\multicolumn{4}{|c|}{$D = 5$}\\
\hline
$(s,t)$ & $(s^\prime,t^\prime)$ & $(p,q)$ & $R$-symmetry \\
\hline
(10,1) & (4,1) & (6,0) & $usp(8)$\\
& (5,0) & (5,1) & $usp(4,4)$\\
\hline
(9,2) & (4,1) &(5,1) & $usp(4,4)$\\
& (3,2) & (6,0) & $sp(8,R)$\\
& (5,0) & (4,2) & $usp(4,4)$\\
\hline
(6,5) & (4,1) & (2,4) & $usp(4,4)$ \\
& (1,4) & (5,1) & $usp(4,4)$\\
& (5,0) & (1,5) & $usp(4,4)$ \\
& (0,5) & (6,0) & $usp(8)$\\
& (3,2) & (3,3) & $sp(8,R)$\\
& (2,3) & (4,2) & $sp(8,R)$\\
\hline
\multicolumn{4}{|c|}{$D = 4$}\\
\hline
(10,1) & (3,1) & (7,0) & $su(8)$ \\
& (4,0) & (6,1) & $su^*(8)$\\
\hline
(9,2) & (3,1) & (6,1) & $su(4,4)$\\
& (4,0) & (5,2) & $su^*(8)$\\
& (2,2) & (7,0) & $sl(8,R)$\\
\hline
(6,5) & (3,1) & (3,4) & $su(4,4)$\\
& (1,3) & (5,2) & $su(4,4)$\\
& (4,0) & (2,5) & $su^*(8)$\\
& (0,4) & (6,1) & $su^*(8)$\\
& (2,2) & (4,3) & $sl(8,R)$\\
\hline
\multicolumn{4}{|c|}{$D =3$}\\
\hline
(10,1) & (2,1) & (8,0) & $so(16)$\\
& (3,0) & (7,1) & $so^*(16)$\\
\hline
(9,2) & (2,1) & (7,1) & $so(8,8)$ \\
&(1,2) & (8,0) & $so(8,8)$\\
& (3,0) & (6,2) & $so^*(16)$\\
\hline
(6,5) & (2,1) & (4,4) & $so(8,8)$ \\
& (1,2) & (5,3) & $so(8,8)$\\
& (3,0) & (3,5) & $so^*(16)$\\
& (0,3) & (6,2) & $so^*(16)$\\
\hline

\end{tabular}
\caption{$R$-symmetries of $11D$ supergravity compactified on Lorentzian tori}
\end{center}
\end{table}

Note that all real forms of $C_4, A_7$ and $D_8$ are a maximal subgroup of the
$U$-duality groups $E_{6(6)}, E_{7(7)}, E_{8(8)}$.  The associated moduli spaces of
these theories are therefore $E_{11-D(11-D)}/H_D$, when $H_D$ are the appropriate
real forms of the $R$-symmetries in Table~8.

\section{State counting and ghosts in $M, M^*, M^\prime$ theories and IIB$^*$
theories}

In this section, we count the degrees of freedom and show that in any theory, with
any signature, supersymmetry implies that there are at most three distinct phases,
two Minkowskian (at least one time), where all 128 bosons either are no-ghost or
they split in 64$^+$ positive norm and 64$^-$ negative norm states.  The other phase
is Euclidean (no time) and the states arrange in a 72$^+$ positive norm and 56$^-$
negative norm sector.  For the Euclidean theories, care is needed to give a
meaning to a state and its norm since there are no truly massless particle in this case. 
Our analysis is made as counting scalar degrees of freedom with their kinetic term
factor after all these theories are dimensionally reduced to three dimensions.  In
this extreme situation, there are only three possibilities, as shown in Table~8,
since the 128 bosons are coordinates of the $\sigma$-models $E_{8(8)}/SO(16)$ or
$E_{8(8)}/SO(8,8)$ for Minkowskian phases or $E_{8(8)}/SO^*(16)$ for the Euclidean
phase.  Note that no other possibility is allowed since there are no other real
forms of $D_8$ contained in $E_{8(8)}$ \cite{gil}.

The first case has no negative norm states since $SO(16)$ is the maximal
compact subgroup of $E_{8(8)}$.  For the other two cases, we precisely get $n^-
= 64, 56$.  

The rest of this section is devoted to showing how these states are arranged with
the spin degrees of freedom when these theories are lifted to higher dimensions.

To do state counting in Minkowskian spaces with arbitrary signatures, one must
use some rules, which are dictated by the underlying gauge invariance of the
higher-dimensional supergravity with respect to coordinate transformations and
gauge transformations of the $p$-form gauge potentials.  For the metric part,
one uses the fact that a massless graviton, in a Minkowskian space $V_{(s,t)}$,
is associated to the coset $SL(s + t -2)/SO(s-1,t-1)$.  The number of positive
and negative norm states is

\begin{eqnarray*}
n^+ &=& \frac{(s + t - 2)(s + t - 1)}{2} - (s - 1)(t-1) - 1 \\ 
n^- &= &\frac{(s + t - 2)(s + t - 3)}{2} - \frac{(s-1)(s-2) + (t-1)(t-2)}{2}~.
\end{eqnarray*}
Similar formulae exist for $p$-forms gauge fields.  The above analysis does not
apply for the Euclidean case $(t = 0)$.  However, in this case, one can use the
rule that a positive norm $3d$ Euclidean vector is dual to a negative norm
Euclidean scalar.  By lifting this rule, using the $KK$ ansatz, one is led to
conclude that a Euclidean graviton in $D$ dimensions parametrizes the coset
$SL(d-2,R)/SO(D-3,1)$.  Curiously, this is identical to the coset of a graviton
in $V_{D-2,2}$ Minkowski space.  The reason for this is that, after
compactification on $T_{D-4}$, one gets $4d$ gravity in (4,0) and (2,2)
space-time and they both give, upon $S_1$ reduction, the $\sigma$-model
$SL(2,R)/SO(1,1)$, where the role of the two scalars has just been interchanged. 
Note that for (3,1) gravity, the $S^1$ reduction gives the standard
$SL(2,R)/SO(2)$ coset.  Another explanation of this rule, related to $M^*$
theory, is given later.  The above rules are also consistent with the fact that
upon reduction on a time-like $S^1$, gravity gives a $KK$ vector with a
reversed sign of the metric (and a scalar with the correct sign), while a
$p$-form gives a $(p-1)$ form with a wrong sign of the metric (in addition to a
$p$-form with the original sign) \cite{cj}.

Let us now consider $M, M^*$ and $M^\prime$ theories.  On an $11D$ Minkowskian
background, the state counting goes as follows:
\begin{eqnarray*}
M~{\rm theory} \quad\quad 128 &=& 44^+ + 84^+ \\
M^*~{\rm theory} \quad\quad 128 &=& 64^+ + 64^-\\
44 &=& 36^+ + 8^-\\
84 & = & 28^+ + 56^- \\
M^\prime~{\rm theory} \quad\quad 128 &=& 64^+ + 64^-\\
44 & = & 24^+ + 20^-\\
84 & = & 40^+ + 44^-
\end{eqnarray*} 

If we reduce $M^*$ theory on $S^1$, we get IIA$^*$ theory and the negative norm
states correspond to the $RR$ vector and three forms.  In a similar way, if we
consider IIB$^*$ theory, the $NS$ and $RR$ states have reversed signs of
the metric and since they are equal in number, they give $128 = 64^+_{NS} +
64^-_{RR}$ on a (9,1) background.  For the Euclidean gravity, IIA$_E$, the
counting can be obtained by time-reduction on $S^1$ from $M$ theory and using
the previous rules, the ghost counting goes as follows:
\begin{eqnarray*}
128 & = & 72^+ + 56^- \\
44 & = & 30^+ + 14^-\\
84 & = & 42 + 42^- 
\end{eqnarray*}
Note that in type IIA$_E$, the $RR$ vector and the $NS$ antisymmetric tensor
have reversed sign, not the $RR$ three-form.
If we instead consider $M^*$ theory as a space-like $S^1$, we get an (8,2) theory
where the $RR$ vector and $RR$ three-forms have reversed sign with respect to
the IIA$_E$ theory.  In the (8,2) theory, these states contribute as
$(36^-,28^+)$ to the total $(64^+,64^-)$ states.  If we flip the signs of these
states, we get $(36^+,28^-)$ added to $(36^+,28^-)$, explaining the $(72^+,56^-)$
signs in the Euclidean case.
In three dimensions, each phase corresponds to a different $\sigma$-model and
there is a one-to-one correspondence.  For higher dimensions, many different
theories can be in the same phase.  For instance in $D = 4,~(64^+,64^-)$
correspond to  two $N = 8$ supergravities with $\sigma$-models $E_{7(7)}/SU(4,4)$ and $E_{7(7)}/SL(8,R)$,
corresponding to (3,1) signature and (2,2) signature, respectively.  The
$(72^+,56^-)$ phase corresponds to $N = 8$ Euclidean supergravity with $\sigma$-model
$E_{7(7)}/SU^*(8)$ \cite{cj}.  The higher the dimension, the more theories correspond to
the same phase.

\subsection{Theories with 16 supersymmetries}

The previous analysis can be extended to any theory with a lower number of
supersymmetries.  Here we just consider theories with 16 supersymmetries.  In $D
= 10$ the (1,0) algebra can be constructed for signatures (9,1) and (5,5).  This
algebra admits both a supergravity and a Yang--Mills theory whose Lagrangians are
identical to the standard (9,1) Lagrangians since their space-time is related by
mod~8 periodicity.  No other signatures are possible for real Weyl fermions.

By compactification on Lorentzian tori, one can get several theories in lower
dimensions.  In particular, descending to $D = 4$, one gets $N = 4$
supergravities and $N = 4$ superconformal Yang--Mills theories with space-time, 
with signature (3,1) (2,2) and (4,0) \cite{ws}--\cite{bvn}.  The relevant 
superalgebras are  $SU(2,2/4), SU(2,2/2,2)$ for a (3,1) signature (corresponding to super
Yang--Mills on $T_6$ and $T_{2,4}$ respectively), $SL(4/4)$ for a (2,2) signature
 (corresponding to super-
Yang--Mills on  $T_{(3,3)}$) and $SU^*(4/4)$ for a (4,0) signature (corresponding
 to super-Yang--Mills on $T_{5,1}$).  
In $D = 4$ the moduli space of (1,0) supergravity on
the corresponding tori are:
\begin{eqnarray*}
(3,1) \times (6,0) &\quad\quad &\frac{SO(6,6)}{SU(4) \times SU(4)} \times
\frac{SU(1,1)}{U(1)} \\
(3,1) \times (2,4) &\quad\quad& \frac{SO(6,6)}{SU(2,2)\times SU(2,2)}
\times \frac{SU(1,1)}{U(1)}\\
(4,0) \times (5,1) &\quad\quad& \frac{SO(6,6)}{SU^*(4) \times SU^*(4)} \times
\frac{SU(1,1)}{SO(1,1)}\\
(2,2) \times (3,3) &\quad\quad& \frac{SO(6,6)}{SL(4,R)\times SL(4,R)} \times
\frac{SU(1,1)}{SO(1,1)}
\end{eqnarray*}
Further reducing to $D = 3$, we get three $\sigma$-models:
$$
\frac{SO(8,8)}{SO(8)\times SO(8)}~,~~\frac{SO(8,8)}{SO(4,4)\times SO(4,4)}~,~~
\frac{SO(8,8)}{SO^*(8) \times SO^*(8)}~.
$$
This again leads to three phases for the 64 (bosonic) degrees of freedom.  In
the Minkowskian phases, either $n^- = 0~(n^+ = 64)$ or $n^- = 32~(n^+ = 32)$. In
the Euclidean phase ,$n^- = 24~(n^+ = 40)$.  Lifting the theory to $D = 10$, the
graviton has $35 = 23^+ + 12^-$ degrees of freedom, the antisymmetric
 tensor $28 = 16^+ + 12^-$, and
the dilaton $1^+$.  The graviton states span the $\sigma$-model
$SL(8,R)/SO(6,2)$.  The positive and negative norm states fall, as usual, in the
representation of the maximal compact subgroup $U(4)$. 

{\large \bf Acknowledgements}

Part of this report is based on collaborations with R. D'Auria, M.A. Lled\'o and
R.~Varadarajan.  Enlightening conversations with  Y. Oz, A. van Proeyen and R. Stora are also
acknowledged.
This work has been supported in part by the European Commission RTN network
HPRN-CT-2000-00131 (Laboratori Nazionali di Frascati, INFN) and by the D.O.E. grant
DE-FG03-91ER40662, Task C.
\vfill\eject


\end{document}